\documentclass[12pt]{article}
\usepackage{psfig}

\usepackage[dvips]{color}

\newcommand{\Black}{\color [rgb]{0,0,0}}

\newcommand{\Brown}{\color [rgb]{0.4,0.1,0.1}}

\def\TL{\hfil$\displaystyle{##}$}
\def\TR{$\displaystyle{{}##}$\hfil}

\def\TT{\hbox{##}}
\def\seqalign#1#2{\vcenter{\openup1\jot
  \halign{\strut #1\cr #2 \cr}}}

\def\fixit#1{}

\def\mop#1{\mathop{\rm #1}\nolimits}

\def\tr{\mop{tr}}

\def\overleftrightarrow#1{\vbox{\ialign{##\crcr
     $\leftrightarrow$\crcr\noalign{\kern-0pt\nointerlineskip}
     $\hfil\displaystyle{#1}\hfil$\crcr}}}

\def\lsim{\mathrel{\mathstrut\smash{\ooalign{\raise2.5pt\hbox{$<$}\cr\lower2.5pt\hbox{$\sim$}}}}}
\def\gsim{\mathrel{\mathstrut\smash{\ooalign{\raise2.5pt\hbox{$>$}\cr\lower2.5pt\hbox{$\sim$}}}}}

\def\sqr#1#2{{\vcenter{\vbox{\hrule height.#2pt
         \hbox{\vrule width.#2pt height#1pt \kern#1pt
            \vrule width.#2pt}
         \hrule height.#2pt}}}}
\def\square{\mathop{\mathchoice\sqr56\sqr56\sqr{3.75}4\sqr34\,}\nolimits}

\def\href#1#2{#2}

\def\lbldef#1#2{\expandafter\gdef\csname #1\endcsname {#2}}
\def\eqn#1#2{\lbldef{#1}{(\ref{#1})}%
\begin{equation} #2 \label{#1} \end{equation}}
\def\eqalign#1{\vcenter{\openup1\jot
    \halign{\strut\span\TL & \span\TR\cr #1 \cr
   }}}

\begin{document}
\pagestyle{plain}
\setcounter{page}{1}
\begin{titlepage}

\begin{flushright}
PUPT-2051 \\
hep-th/0210093
\end{flushright}
\vfil

\begin{center}
{\huge Double-trace operators and}
\vskip0.5cm
{\huge one-loop vacuum energy in AdS/CFT}
\end{center}

\vfil
\begin{center}
{\large Steven S. Gubser and Indrajit Mitra}
\end{center}

$$\seqalign{\span\TL & \span\TT}{
& Joseph Henry Laboratories, Princeton University, Princeton, NJ 08544
}$$
\vfil

\begin{center}
{\large Abstract}
\end{center}

\noindent
We perform a one-loop calculation of the vacuum energy of a tachyon
field in anti de-Sitter space with boundary conditions corresponding
to the presence of a double-trace operator in the dual field theory.
Such an operator can lead to a renormalization group flow between two
different conformal field theories related to each other by a Legendre
transformation in the large $N$ limit.  The calculation of the
one-loop vacuum energy enables us to verify the holographic c-theorem
one step beyond the classical supergravity approximation.

\vfil
\begin{flushleft}
October 2002
\end{flushleft}
\end{titlepage}
\newpage
\renewcommand{\thefootnote}{\arabic{footnote}}
\setcounter{footnote}{0}
\tableofcontents

\section{Introduction}
\label{Introduction}

The AdS/CFT correspondence \cite{Malda,GKP,WittenAdS} (for reviews see
\cite{MAGOO,DHokerDan}) relates a $d$-dimensional quantum field theory
to a $(d+1)$-dimensional gravitational theory, the most notable
example being ${\cal N}=4$, $d=4$ super-Yang-Mills theory and type~IIB
string theory on $AdS_5 \times S^5$.  Most of the checks and
predictions of this duality have been at the level of classical
supergravity.  It is particularly difficult to carry out meaningful
loop computations in AdS, corresponding to $1/N$ corrections in the
gauge theory, simply because the supergravity theory is highly
non-renormalizable, and the Ramond-Ramond fields make computations in
the string genus expansion unwieldy at best.  The aim of this note is
to obtain a simple one-loop result in AdS that is finite in any
dimension.  The result is an expression for the difference of the
vacuum energies that arises from changing boundary conditions on a
tachyonic scalar field with mass in a particular range.

The inspiration for this computation came from Witten's treatment
\cite{WittenMulti} of multi-trace deformations of the gauge theory
lagrangian and their dual descriptions in asymptotically anti-de
Sitter space. Such a dual description was also discussed in
\cite{SecMulti}; however, our treatment will follow \cite{WittenMulti}
more closely. Earlier work describing the same gauge theory
deformations in terms of non-local terms in the string worldsheet
action appeared in \cite{ABSMulti,ABSMultiTwo}.  To be definite,
suppose one were to add to the gauge theory lagrangian a term ${f
\over 2} {\cal O}^2$ where ${\cal O}$ is a single trace operator with
dimension $3/2$, dual to a scalar field $\phi$ whose mass satisfies
$m^2 L^2 = -15/4$.\footnote{Such a situation could arise in the theory
dual to D3-branes at the tip of a conifold \cite{kwOne}, where there
are indeed dimension $3/2$ color singlet operators.}  The coefficient
$f$ has dimensions of mass, so ${f \over 2} {\cal O}^2$ is a relevant
deformation, and there is a renormalization group (RG) flow starting
from a UV fixed point where $f=0$.  The endpoint of this flow is,
plausibly, an IR fixed point whose correlators are related to those of
the original $f=0$ theory, in the large $N$ limit, by a Legendre
transformation in a manner explained in \cite{kwTwo}.\footnote{We will
discuss further in section~\ref{Multi} the reasoning behind the claim
that the flow ends at an IR fixed point, as well as some caveats.}  In
particular, the scalar that was for $f=0$ related to the operator
${\cal O}$ of dimension $3/2$, is at the IR fixed point related to an
operator $\tilde{\cal O}$ of dimension $5/2$.

How is all this reflected in AdS?  According to \cite{WittenMulti},
the addition of ${f \over 2} {\cal O}^2$ amounts to specifying
particular linear boundary conditions on the scalar $\phi$ at the
boundary of AdS.  At the classical level, these boundary conditions
are consistent with the original $AdS_5$ solution with $\phi=0$.
Superficially, this looks like a puzzle, since we were expecting an RG
flow.  In fact, conformal invariance is violated by the ${\cal O}^2$
deformation, but at leading order in $N$ its effects are restricted to
certain correlators that we will describe in section~\ref{Multi}.  The
crux of the matter is that it is impossible to satisfy the boundary
conditions on $\phi$ with a $SO(4,2)$-invariant bulk-to-bulk
propagator, except when $f=0$ or $\infty$.  This gives rise to one
loop effects that cause deviations from $AdS_5$.

Although we will not obtain the full one-loop corrected solution
corresponding to an RG flow due to the ${f \over 2} {\cal O}^2$
deformation, we will consider its endpoints and perform a one-loop
supergravity check of the c-theorem.  This ``theorem,'' conjectured in
four dimensions by Cardy \cite{Cardy} as a generalization of
Zamolodchikov's celebrated two-dimensional c-theorem
\cite{Zamolodchikov}, has been shown to follow from AdS/CFT at the
level of classical supergravity provided the null energy condition
holds \cite{gppzOne,fgpwOne} (see also \cite{Alvarez:1998wr} for
earlier work in this direction).  The magnitude of the vacuum energy
of $AdS_5$, measured in five-dimensional Planck units, is proportional
to an appropriate central charge raised to the $-2/3$ power.  So the
vacuum energy should be more negative in the infrared than in the
ultraviolet, and at the classical level, that is what is shown in
\cite{gppzOne,fgpwOne} (actually, the arguments on the AdS are
dimension-independent, though it is not entirely how to translate the
``holographic'' central charge into field theory language in the case
of odd-dimensional CFT's).  At the quantum level, the arguments of
\cite{gppzOne,fgpwOne} have no force because it's not clear that the
null energy condition is valid or even relevant.  So an explicit loop
calculation is appropriate.  All that is needed is the one-loop
contribution of the scalar $\phi$ to the vacuum energy.  This quantity
is divergent, but the difference between imposing the two simple
boundary conditions (described above as $f \to 0$ and $f \to \infty$)
gives a finite result.  The contributions of all other fields can be
ignored because they do not change at the one loop level as one
changes the boundary conditions on $\phi$.  Also, because we only
desire a one-loop vacuum amplitude, we may entirely ignore
interactions of the scalar with other fields, and work simply with the
free action
  \eqn{ScalarAct}{
   S = \int d^5 z \sqrt{g} \, \left( -{1 \over 2} (\partial\phi)^2 -
    {1 \over 2} m^2 \phi^2 \right) \,,
  }
 where we work in mostly plus signature, so that the metric of $AdS_5$
on the Poincar\'e patch is
  \eqn{AdSMetric}{
   ds^2 = {L^2 \over z^2} \left( -dt^2 + d\vec{x}^2 + dz^2 \right) \,.
  }
 For definiteness, our discussion has focused on $AdS_5$ and a scalar
with a particular mass; however, the results we will obtain can be
presented with considerable generality for $AdS_{d+1}$, as we will
describe.  For odd $d$, the formulas for the vacuum energy are much
more complicated, and for the sake of efficiency we check the sign via
numerics.

The organization of the paper is as follows. In section~\ref{Multi} we
briefly review the prescription of \cite{WittenMulti} for treating
multi-trace operators, and we demonstrate that general boundary
conditions are incompatible with $SO(4,2)$-invariance of the scalar
propagator.  In section~\ref{OneLoop} we compute the finite change in
the one-loop vacuum energy discussed above, and make some remarks on
the interpolating geometry connecting the two anti-de Sitter
endpoints.  We conclude in section~\ref{Conclusions} by extracting the
prediction for the central charge, and observing that the c-theorem is
obeyed.

\section{Multi-trace operators and scalar propagators}
\label{Multi}

The proposal of \cite{WittenMulti} is a natural generalization of the
original prescription for computing correlators \cite{GKP,WittenAdS},
and it should in principle be derivable from it: see
\cite{Minces:2002wp} for a more precise discussion.  Suppose one
starts with the complete set ${\cal O}_a$ of independent, local,
color-singlet, normalized, single-trace operators: for ${\cal N}=4$
super-Yang-Mills theory these would include, for example, ${1 \over N}
\tr X_1 X_2$ and ${1 \over N} \tr F_{\mu\nu} \nabla_\rho \lambda_1$.
The action can be written as $I = N^2 W({\cal O}_a)$ for some
functional $W$, which for ${\cal N}=4$ super-Yang-Mills would be the
integral of a linear function of those ${\cal O}_a$ which are Lorentz
scalars.  The general belief is that the ${\cal O}_a$ can be put into
one-to-one correspondence with the quantum states of type~IIB string
theory in $AdS_5$.\footnote{There is considerable subtlety in this
claim. It has
been demonstrated that the Kaluza-Klein tower of supergravity modes in
$AdS_5 \times S^5$ is in correspondence with the chiral primaries of
${\cal N}=4$ super-Yang-Mills and their descendants; and the duals of
certain non-perturbative states have been found, such as dibaryons
(see for example \cite{gkBaryon} and giant gravitons \cite{McGreevy:2000cw}.
Evidence is growing that the operator-state map extends faithfully to
excited string states (see for example \cite{bmn,gkPolTwo}).  Since
the states in question can sometimes be extended across most of
$AdS_5$ (as in \cite{gkPolTwo}), it is not entirely clear that a
second quantized treatment in terms of local fields is appropriate;
but this is scarcely relevant to the situation at hand, since extended
states are very massive, and we're interested only in tachyons.}
Restricting ourselves to scalars in $AdS_5$, we have the standard
relation $\Delta_a (\Delta_a - d) = m_a^2 L^2$ relating the dimension of
${\cal O}_a$ to the mass of the field $\phi_a$.  Writing the metric
for the Poincar\'e patch of $AdS_5$ as
  \eqn{Poincare}{
   ds^2 = {L^2 \over r^2} \left( -dt^2 + \sum_{i=0}^{d-2} dx_i^2 +
    dr^2 \right) \,,
  }
 we have boundary asymptotics for $\phi_a$ as follows:
  \eqn{boundary}{
   \phi_a \sim \alpha_a(x) r^{d - \Delta_a} + \beta_a(x) r^{\Delta_a} \qquad
    \hbox{for $r \to 0$.}
  }
 The prescription of \cite{WittenMulti} is to replace $W({\cal O}_a)$
by $W(\beta_a)$ and impose the following boundary conditions:
  \eqn{Wboundary}{
   \alpha_a(x) = {\delta W \over \delta \beta_a(x)} \,.
  }
 The partition function of the gravitational theory in AdS, subject to
the boundary conditions \Wboundary, is then supposed to equal the
partition function of the gauge theory.

The simplest non-trivial example is double trace operators: most
simply, ${\cal O}^2$ where the scalar operator ${\cal O}$ has
dimension $\Delta$ between ${d \over 2}-1$ and $d/2$.  Precisely in
this range, unitarity bounds are satisfied, and both power law
behaviors in \boundary\ are normalizable.  Then $W$ includes a term
${f \over 2} \int d^d x \, {\cal O}^2$.  This brings us back to the
discussion initiated in the introduction: nonzero $f$ plausibly drives
the field theory from a UV fixed point where the boundary conditions
are $\alpha=0$ to an IR fixed point where the boundary conditions are
$\beta=0$.  Since these two fixed points will be the focus of
section~\ref{OneLoop}, let us introduce an additional convenient
notation: $\Delta_+$ and $\Delta_-$ are the two solutions to $\Delta
(\Delta - d) = m^2 L^2$, with $\Delta_-$ being the lesser of the two
(and thus in the aforementioned range, from ${d \over 2}-1$ to $d/2$).
Clearly $\Delta_+ = d-\Delta_-$.

When $\Delta_- < d/2$, the addition of a trace-squared operator ${\cal
O}^2$, where ${\cal O}$ has dimension $\Delta=\Delta_-$, is a relevant
deformation, so conformal invariance must be broken in the gauge
theory.  The results of \cite{WittenMulti} for $d=4$ and $\Delta_-=2$
suggests that even when $\Delta_- = d/2$ there is a logarithmic RG
flow.  The simplest indication of the breaking of conformal invariance
in supergravity is that the bulk-to-bulk propagator for the scalar
$\phi$ dual to ${\cal O}$ cannot be $SO(4,2)$-invariant.  We will now
demonstrate this claim.

The propagator in question can be defined as
  \eqn{DefProp}{
   i G(z,z') = \langle 0 | T\{ \phi(z) \phi(z') \} | 0 \rangle \,,
  }
 and it satisfies the equation of motion
  \eqn{PropEOM}{
   (\square-m^2) G(z,z') = \delta^{d+1}(z-z') \,,
  }
where $\square = g^{\mu\nu}\nabla_\mu\nabla_\nu$, and the delta
function includes a $1/\sqrt{g}$ in its definition, so that
  \eqn{DeltaDef}{
   \int d^{d+1}z \sqrt{g} \, f(z) \delta(z-z') = f(z')
  }
 for any continuous function $f(z)$.  If the propagator is to respect
$SO(4,2)$ invariance, it must be a function only of the geodesic
distance $\sigma(z,z')$, which is known to be
  \eqn{ellZetaDef}{
   \sigma(z,z') = L \log\left( {1 + \sqrt{1-\zeta^2} \over \zeta} \right)
    \qquad\hbox{where}\qquad
   \zeta = {2rr' \over r^2 + r'^2 - (t - t')^2 + (\vec{x}-\vec{x}')^2} \,,
  } 
where $L$ is the radius of AdS.  The only solutions to \DeltaDef\
which are functions only of $\zeta$ are $G(z,z') = p G_{\Delta_-} +
(1-p) G_{\Delta_+}$ where for any $\Delta$
(cf.~\cite{Fronsdal,BurgLut}),\footnote{The expression for $G(z,z')$
above differs by a sign from that in \cite{BurgLut,Fronsdal} because
the latter define the Green's function as $- i G(z,z') = \langle 0|
{\rm T} \phi (z) \phi (z') |0 \rangle$.}
  \eqn{GreensFunct}{
   i G_\Delta = {{\Gamma(\Delta)} \over {2^{\Delta} \pi^{d/2}L^{d-1} (2
    \Delta - d) \Gamma(\Delta - {d \over 2})}} \zeta^{\Delta}
    F\left({\Delta
    \over 2}, {{\Delta + 1} \over 2}; \Delta - {d \over 2} + 1;
    \zeta^2 \right) \,.
  }
By keeping $z'$ fixed while $z$ approaches the boundary of AdS, it is
straightforward to verify that for no choice of $p \in (0,1)$ and $f
\in (0,\infty)$ does the propagator $G(z,z') = p G_{\Delta_-} + (1-p)
G_{\Delta_+}$ satisfy the boundary conditions \Wboundary, which in our
case amount to $\alpha = f\beta$.  For $p=0$ and $f=0$ the boundary
conditions are satisfied with $SO(4,2)$ invariance preserved,
corresponding to a fixed point of RG where $\phi$ is dual to an
operator ${\cal O}$ with dimension $\Delta_-$.  Let us call this the
$\Delta_-$ theory.  And for $p=1$ and $f=\infty$ (formally speaking),
again the boundary conditions are satisfied with $SO(4,2)$ invariance,
and now $\phi$ corresponds to an operator $\tilde{\cal O}$ with
dimension $\Delta_+$: this we will call the $\Delta_+$ theory.

It was already remarked in \cite{WittenMulti} that a renormalization
group flow should interpolate between the $\Delta_-$ theory in the UV
and the $\Delta_+$ theory in the IR.  This is in fact a somewhat
subtle claim: why should we think that the RG flow initiated by adding
${f \over 2} {\cal O}^2$ ends up at a non-trivial IR fixed point?  We
can argue as follows:\footnote{SSG thanks E.~Silverstein for a
discussion in which the following line of reasoning arose.} the
Legendre transformation prescription of \cite{kwTwo} guarantees that
the IR fixed point exists, at least in the large $N$ limit.  The
existence of a fixed point of RG is a generic phenomenon, so $1/N$
corrections should not spoil the claim, nor should they greatly alter
the location of the fixed point in the space of possible couplings.
Since a naive scaling argument (just looking at the dimension of $f$)
tells us that the RG flow should end up at the desired IR fixed point
if we ignore all $1/N$ corrections, it should be that {\it some} RG
flow exists close to the approximately one we naively identified,
ending at the non-trivial IR fixed point.  A significant caveat to
this reasoning is that AdS/CFT examples often (in fact, nearly always
in the literature so far) have exactly marginal deformations.  A {\it
line} of fixed points of RG is {\it not} a generic phenomenon, and
$1/N$ effects in the absence of supersymmetry generically could
destroy such a line.  Only one point could be left after $1/N$ effects
are included; or, worse yet, only a point infinitely far out in
coupling space could be left.  Translated into supergravity terms,
these remarks mean that the one-loop contribution to the potential
could source the dilaton or other moduli, possibly leaving no extrema
at finite values of the fields.  If there are no such moduli in the
first place (as perhaps one would expect for a truly {\it generic}
non-supersymmetric quantum field theory with an AdS dual), then this
caveat is not a problem.  In practice, however, it is likely to
interfere with constructing explicit string theory examples of the RG
flow discussed in this paper.  For the remainder of our discussion, we
will ignore the caveat.

Since the renormalization group flow is non-trivial, it is natural to
expect that the supergravity geometry deviates from AdS.  The surprise
is that this does {\it not} happen classically.  Roughly, this can be
understood in field theory terms as a reflection of the fact that
$n$-point functions involving only the stress energy tensor do not
receive corrections at leading order in $N$.\footnote{Correlation
functions which {\it do} receive corrections at leading order in $N$
when ${f \over 2} {\cal O}^2$ is added to the lagrangian are precisely
those which pick up contributions from factorized forms $\langle {\cal
O} \ldots \rangle \langle {\cal O} \ldots \rangle$, where the dots
indicate any arrangement of the operators involved in the original
correlator.}  At subleading order in $N$, or at one-loop in
supergravity, deviations from AdS must occur, simply because a
one-loop diagram where the $SO(4,2)$-non-invariant scalar propagator
closes upon itself must give rise to an effective potential that
varies over spacetime.  Entertainingly, there is no classical scalar
field which is varying; rather, the variation in the potential arises
on account of proximity to the boundary.  This is in contrast to
previously studied examples of RG flow in $AdS_5$ (for instance
\cite{gppzOne,fgpwOne}) where the flow is described in terms of
scalars in the five-dimensional supermultiplet of the graviton with
non-trivial dependence on radius.

There should be a solution to the one-loop-corrected supergravity
lagrangian interpolating between one asymptotically AdS region near
the boundary, corresponding to the $\Delta_-$ UV fixed point, and a
different one in the interior, corresponding to the $\Delta_+$ IR
fixed point.  For instance, one could require that the symmetries of
${\bf R}^{3,1}$ be preserved in the solution, which must then have the
form
  \eqn{RGForm}{
   ds^2 = e^{2A(r)} (-dt^2 + d\vec{x}^2) + dr^2 \,,
  } 
where $A(r) \to r/L_\mp$ as $r \to \pm\infty$.  (Another choice would
be to require the symmetries of ${\bf S}^3 \times {\bf R}$, which
should lead to a solution with the conformal structure of global
AdS).  We will not find the full interpolating solution, but we will
explore some properties of its AdS endpoints.  We will be
particularly interested in the central charge of the CFT's dual to the
two endpoints.  To the leading non-trivial order, these may be
computed as a one-loop saddle-point approximation to the supergravity
``path integral'' (supposing that such an object exists), but without
deforming the AdS background itself.

\section{One loop vacuum energy for the tachyon field}
\label{OneLoop}

The full classical action that we wish to consider is
 \eqn{FullAction}{
  S = {1 \over 2\kappa^2} \int d^{d+1}z \sqrt{g} \, (R - \Lambda_0) + 
   \int d^{d+1}z \sqrt{g} \left( -{1 \over 2} (\partial\phi)^2 - 
    {1 \over 2} m^2 \phi^2 \right) \,.
  }
Here $\Lambda_0$ is a negative constant.  The scalar is subject to the
boundary conditions
 \eqn{LinearBoundaryConditions}{
  \phi \sim \alpha r^{d - \Delta} + \beta r^\Delta \qquad
   \hbox{where} \qquad \alpha = f\beta \,.
 }
As remarked previously, $AdS_{d+1}$ with $\phi=0$ and $1/L^2 = -
{\Lambda_0 \over d(d-1)}$ is a classical solution to the equations of
motion from \LinearBoundaryConditions, but we expect that once
one-loop effects are accounted for, this solution is corrected to an
interpolation between $AdS_{d+1}$ spaces in the UR and IR with
slightly different radii.  The one-loop scalar bubble diagram corrects
the gravitational lagrangian by an amount $\delta {\cal L}$, where
 \eqn{LambdaCorrected}{
  -\sqrt{g}^{-1} \delta {\cal L} = V = -{i \over 2} \tr\log(-\square+m^2) \,.
 } Our main computation will be to evaluate this correction in the
unperturbed background.  In principle, one could go on to find the
interpolating geometry perturbatively in the small parameter $\kappa
\Lambda_0^{(d-1)/2}$.  This would require separating $\delta {\cal L}$
into contributions to the cosmological term and two- and
four-derivative expression in the metric---a much more involved
computation than simply evaluating \LambdaCorrected\ in the
unperturbed background.  For brevity, we will use the notation $V$ in
preference to $\delta {\cal L}$ for the scalar self-energy
\LambdaCorrected, despite the fact that in the full
background-independent form involves derivative terms as well as
finite non-local terms.  $V$ is divergent, but we assume that the
action \FullAction\ is part of well-defined theory of quantum gravity
(presumably, a compactification of string theory or M-theory), so that
all loop divergences are canceled in some physical way, leaving only
finite renormalization effects.  It may be that in the full theory,
$\Lambda_0$ is just the extremal value of a classical potential
function of several scalars; if so, then we are operating on the
understanding that the second derivative of this potential function
with respect to $\phi$ vanishes at $\phi=0$ (that is, we've soaked up
any such second derivative into what we call $m^2$ in \FullAction).

In general, it is difficult to compute one-loop corrections in an
effective theory without knowing precisely how the full theory cancels
divergences.  Results obtained for a chiral anomaly in supergravity
\cite{Bilal:1999ph} for $AdS_5 \times S^5$ can be used to show that
the central charge is corrected at one loop in supergravity, leading
to $c \propto N^2-1$, as appropriate for $SU(N)$ super-Yang-Mills,
rather than $c \propto N^2$ (the leading order result).  Thus in this
case, the difficulties were overcome.  Our situation is more generic,
in that we do not depend on supersymmetry or a special spectrum of
operators.  What we are nevertheless able to do is to determine the
finite difference between $V$ in the case where $f=0$ in
\LinearBoundaryConditions\ and the case where $f=\infty$.  This we
will then translate into a change in the central charge as one flows
from the UV (the $\Delta_-$ theory) to the IR (the $\Delta_+$ theory).
What makes the computation clean is that at one loop, we do not have
to worry about interactions of the scalar with other fields, and the
only relevant diagram is the one where a single scalar propagator
closes on itself, with no vertices.

\subsection{Vacuum energy in limiting regions of AdS}
\label{Endpoints}

The computation of the one-loop contribution to the vacuum energy by a
scalar in curved space, like in flat space, amounts to summing the
logarithm of the eigenvalues of the Klein-Gordon operator. A more
easily computable expression is obtained by expressing the result in
terms of an integral of the Green's function with respect to some
parameter such as proper time or mass.\footnote{For a different method
of computing the effective potential based on the technique of
Zeta-function regularization see \cite{Caldar}.}  All of this is quite
standard, so we just write down the result, referring the reader to
\cite{BDBook} pp.~156-158 for a derivation: if the propagator
$G(z,z';m^2,f)$ is defined by
 \eqn{RepGEOM}{
   (\square_z-m^2) G(z,z';m^2,f) = \delta^{d+1}(z-z') \,,
  } 
(with the delta-function including a $\sqrt{g}$ factor as in
\DeltaDef) together with boundary conditions
\LinearBoundaryConditions, as discussed in section~\ref{Multi}, then
formally,
 \eqn{IntV}{
  V(z;m^2,f) = -{i \over 2} \lim_{z\to z'} 
   \int_{m^2}^\infty d\tilde{m}^2 \, 
   G(z,z';\tilde{m}^2,f) \,,
 }
and for the cases $f=0,\infty$, the fact that we can make the scalar
propagator $SO(4,2)$ invariant means that $V$ will be independent of
the position $z$.\footnote{Actually, we have tucked an additional
complication into our notation: $V$ is, more properly, minus the
one-loop correction to the full gravitational lagrangian, and as such
includes not just a scalar piece, but also terms depending on
curvatures.  For the central charge computation, as we shall explain,
the relevant quantity is the sum of all these terms evaluated on
AdS.}  The formula \IntV\ is problematic because for large masses,
$G(z,z',\tilde{m}^2,0)$ diverges at the boundary of AdS.  This is
unusual: the typical situation for quantum field theory in curved
spacetime is that quantities become well-defined in the limit where
masses are much larger that the inverse radius of curvature.  Thus,
instead of using \IntV, a well-defined procedure is to integrate down
to the Breitenlohner-Freedman bound which is the smallest mass
possible with normalizable modes in AdS.  Thus we obtain
 \eqn{ReplaceV}{
  V(z;m^2,f) = V(z;m^2_{BF},f) + {i \over 2} \lim_{z\to z'} 
   \int_{m^2_{BF}}^{m^2} d\tilde{m}^2 \, G(z,z';\tilde{m}^2,f) \,,
 }
where $m^2_{BF} L^2 = -d^2/4$ is the Breitenlohner-Freedman bound. (For a derivation see the Appendix). It
is possible to argue that $V(z;m^2_{BF},f)$ is the same for $f=0$ and
$f=\infty$.  Indeed, the eigenmodes for a tachyon of mass $m^2$ with
boundary conditions specified by $f=0$ is given by
$\omega = \Delta_- + \ell + 2n$ and that specified by $f = \infty$ is given by $\omega = \Delta_+ + \ell + 2n$ \cite{BF}, where $\ell$ is the
orbital angular momentum quantum number and $n$ is the radial quantum
number. But for a scalar with mass saturating the BF bound, $\Delta_+
= \Delta_- = {d \over 2}$.  So from a viewpoint of canonical
quantization it seems inevitable that $V(z;m^2_{BF},0) -
V(z;m^2_{BF},\infty) = 0$. We can argue further that for general $f$
the eigenfunctions would be a linear combination of those with $f=0$
and $f= \infty$. That would again imply that for $\Delta = {d \over 2}$,
the eigenvalues are unchanged. So we conclude that the $V(z;m_{BF}^2,f) - V(z;m_{BF}^2,0) = 0$ for all values of $f$. 

Thus we are led to the formula that we will really use for
computation:
 \eqn{FinalVDiff}{
  V_+ - V_- = {i \over 2} \int_{m_{BF}^2}^{m^2} d\tilde{m}^2 \,
   \left[ G_{\tilde{\Delta}_+}(z,z)-G_{\tilde{\Delta}_-}(z,z) \right] + 
   V(z;m^2_{BF},\infty) - V(z;m^2_{BF},0) \,,
 }
where $V_+ = V(z,m^2,\infty)$ and $V_- = V(z,m^2,0)$.  We have used
the fact that $G_{\tilde{\Delta}_+}(z,z')$, as defined in
\GreensFunct, is precisely $G(z,z';\tilde{m}^2,\infty)$, while
$G_{\tilde{\Delta}_-}(z,z') = G(z,z,';\tilde{m}^2,0)$.  In light of
the argument of the previous paragraph, the terms outside the integral
cancel.  The advantage of \FinalVDiff\ is that
$G_{\tilde{\Delta}_+}(z,z)-G_{\tilde{\Delta}_-}(z,z)$ is finite, so
that the final answer is also manifestly finite.  We have confidence
that no other finite renormalization effects can slip in to the
calculation, because the only thing that changes between the
$\Delta_-$ and $\Delta_+$ vacua is the boundary condition on $\phi$.

As a warm-up let us first carry out the computation for $AdS_5$. To
get the value of $G_{{\tilde \Delta}_+}(z,z) - G_{{\tilde \Delta_-}}(z,z)$ for
coincident points one has to first express the Green's functions in
terms of the geodesic distance $\sigma$. From (\ref{ellZetaDef}) we
see that in terms of the variable $\zeta$ the geodesic separation is
given by $\cosh({\sigma \over L}) = {1 \over \zeta}$ so we rewrite the
propagator \GreensFunct\ in terms of $\sigma$ and then expand
$i \left[G_{{\tilde \Delta}_+}(z,z) - G_{{\tilde \Delta}_-}(z,z) \right]$ in a power series
in powers of ${\sigma \over L}$. The answer is finite and in the limit
${\sigma \over L} \to 0$, for $AdS_5$ we obtain the simple
expression:
\eqn{GDiffOdd} { 
 i \left[G_{{\tilde \Delta}_+}(z,z) - G_{{\tilde \Delta}_-}(z,z) \right] = - i \left[ G_{{\tilde \Delta}}(z,z) - G_{4 - {\tilde \Delta}}(z,z) \right] = - {{({\tilde \Delta} - 1)({\tilde \Delta} - 2)({\tilde \Delta} - 3)} \over {12 \pi^2 L^3}} \,.
}
The difference in the vacuum energies using (\ref{FinalVDiff}) is therefore
 \eqn{VOdddiff}{\eqalign{
   V_+ - V_- &= {i \over 2} \int_{{m_{\rm BF}}^2}^{m_0^2} d {\tilde m}^2 \left[
   G_{{\tilde \Delta}_+}(z,z) - G_{{\tilde \Delta}_-}(z,z) \right]  \, \cr
   &= - { 1 \over 2} \int^{\Delta_-}_{2} { d {\tilde \Delta} \over L^2} \left[ 2({\tilde \Delta} - 2){{({\tilde \Delta} - 1)({\tilde \Delta} - 2)({\tilde \Delta} - 3)} \over {12 \pi^2 L^3}} \right] \, \cr
   &= - {1 \over {12 \pi^2 L^5}} \int^{\Delta_- - 2}_{0} d {\tilde \nu} \left[{\tilde \nu}^2 ({\tilde \nu}^2 - 1) \right] = {1 \over {12 \pi^2 L^5}} \left[{{(\Delta_- - 2)^3 \over 3}} - {{(\Delta_- - 2)^5} \over 5}  \right] \,,
 }}
where in the second line we have used ${\tilde m}^2 L^2 = {\tilde \Delta}
({\tilde \Delta} - 4)$ and the fact that $\Delta_{BF} = {d \over 2} =
2$. Since $\Delta_- < 2$ we find that $V_+ - V_- < 0$, and therefore
$c_- > c_+$ in agreement with the field theory prediction.

It is straightforward to generalize this for any odd-dimensional anti
de-Sitter spacetime because for $d$
even, the difference $i [G_{{\tilde \Delta}_+}(z,z) - G_{{\tilde \Delta}_-}(z,z) ]$ is quite simple in
form. Before writing this down, for convenience, let us define $d
\equiv 2k$ so that the spacetime is $AdS_{2k+1}$. In terms of $k$,
$i \left[G_{{\tilde \Delta}_+}(z,z) - G_{{\tilde \Delta}_-}(z,z) \right]$ is: 
\eqn{DiffGOdd} { 
 i \left[ G_{{\tilde \Delta}_+}(z,z) - G_{{\tilde \Delta}_-}(z,z) \right] = - i \left[G_{{\tilde \Delta}}(z,z) - G_{d - {\tilde \Delta}}(z,z) \right] = - {{(-1)^{k}} \over {n_k \pi^k L^{d-1}}} \prod_{i=1}^{2k -
1}
   ({\tilde \Delta} - i) \,,
}
where $n_k = 2^k (2k-1)!!$.

The difference in the vacuum energies is therefore
\eqn{OddVdiff}{\eqalign{
 V_+ - V_- &= {i \over 2} \int_{{m_{\rm BF}}^2}^{m_0^2} d {\tilde m}^2 \left[
G_{{\tilde \Delta}_+}(z,z) - G_{{\tilde \Delta}_-}(z,z) \right]  \, \cr
 &= { 1 \over 2} \int_{\Delta_-}^{k} { d {\tilde \Delta} \over L^2} \left[ 2({\tilde \Delta} - k)
{{(-1)^k} \over {n_k \pi^k L^{d-1}}} \prod_{i=1}^{2k - 1} ({\tilde \Delta} - i)
\right] \,,
}}
where in the second line we have used ${\tilde m}^2 L^2 = {\tilde \Delta} ({\tilde \Delta} - d)$ and the fact that $\Delta_{BF}
= {d \over 2} = k$. Shifting the variable of integration by introducing
a new variable ${\tilde \nu} \equiv {\tilde \Delta}_- - k$, the integrand can be written
down in a terms of the Pochhammer symbol $(a)_n = {{\Gamma(a+n)} \over
{\Gamma(n)}}$:
\eqn{Prod}{
  2({\tilde \Delta} - k) {{(-1)^k} \over {n_k \pi^k L^{d+1}}} \prod_{i=1}^{2k - 1}
({\tilde \Delta} - i) =  {{(-1)^k} \over {n_k \pi^k L^{d+1}}} \prod_{i=0}^{k-1}
\left( {\tilde \nu}^2 - i^2 \right) = {1 \over {n_k \pi^k L^{d+1}}} ({\tilde \nu})_k (- {\tilde \nu})_k \,.
}
The factor $(-1)^k$ was nullified by an extra factor of $(-1)^k$ from
the product. Assembling all of this, we finally have

 \eqn{FinalDiffV}{
  V_+ - V_- = {1 \over {2 n_k \pi^k L^{d+1}}} \int_{\nu}^{0} d {\tilde \nu}
   \left[ ({\tilde \nu})_k (-{\tilde \nu})_k \right] \,,
 }
where we recall that $n_k = 2^k (2k-1)!!$. The lower limit of
integration $\nu$ depends on the value of $\Delta_-$. Since $k \leq
\Delta_- \leq k-1 $, the range of $\nu$ is $-1 \leq \nu \leq 0$. The
function $(\nu)_k (-\nu)_k < 0$ for all $k$ and $-1 \leq \nu \leq
0$. So for any odd-dimension anti de-Sitter spacetimes we have shown
that $V_+ - V_- < 0$.

\begin{figure}
   \centerline{\psfig{figure=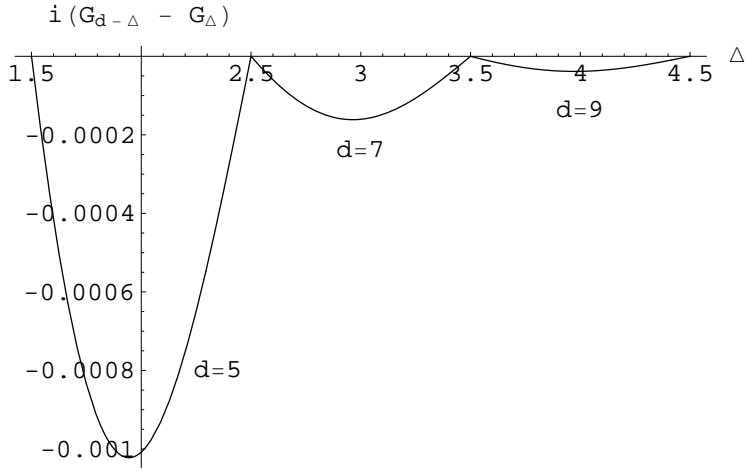,width=4in}}
   \caption{$i (G_{d-\Delta}-G_\Delta)$ as a function of $\Delta$, for
$AdS_6$, $AdS_8$, and $AdS_{10}$ (corresponding to $d=5,7,9$), in
units where $L=1$.}\label{figA}
  \end{figure}

For even dimensional spacetimes, an analytic proof seems cumbersome,
so we resorted to numerics.  As an explicit example,
figure~\ref{figA}
shows a plot of $i (G_{d- \Delta} - G_{\Delta})$ as a function of
$\Delta$ for several even-dimensional anti-de Sitter spacetimes.  In
each dimension, we've plotted the integrand of (\ref{FinalVDiff}) for
$d/2-1 < \Delta < d/2$.  Since the integrand is always negative on
this range, we conclude that $V_+ < V_-$ in accordance with the
c-theorem intuition.  This is also true for $d = 3$, and we believe it
is true generally.

\subsection{Vacuum energy throughout AdS}
\label{Throughout}

The results of the previous section were stated in terms of $V_+ - V_-
= V(z;m^2,\infty) - V(z;m^2,0)$ (both terms were in fact independent
of the position $z$ in AdS).  Here we would like to investigate
$V(z;m^2,f)$ for finite $f$.  This quantity diverges, but
$V(z;m^2,f)-V_-$ is finite.  We will be able to verify the formulas
  \eqn{Vlimits}{
   \lim_{z_0 \to 0} \left[ V(z;m^2,f) - V_- \right] = 0 \,, \qquad
   \lim_{z_0 \to \infty} \left[ V(z;m^2,f) - V_- \right] = V_+ - V_- \,,
  }
which we consider intuitively obvious since ${f \over 2} {\cal O}^2$
is a relevant operator in the CFT, and therefore unimportant in the UV
but important in the IR.

As a first step, one needs the Green's function for the scalar obeying
mixed boundary conditions for all values of $f$ (not just the ones for
$f=0$ and $f= \infty$ that we wrote down earlier). This would be
needed to compute the vacuum energy contribution due to the bubble
diagram. The one-loop corrected action would then induce corrections
in the geometry which can be computed from the Einstein equations. Let us work in Euclidean $AdS$ to get the Green's function $G_E (x,y;f)$ which we shall Wick rotate to obtain $G(x,y;f)$ in Minkowski signature. We
shall follow the canonical method of obtaining Green's functions. In
Poincar\'e coordinates the scalar wave equation is 
 \eqn{wave}{ 
  \left[ x_0^2 (\vec{\partial}^2 + \partial_0^2) - x_0 (d-1) \partial_0 - m^2 \right]
    \phi(x_0, \vec{x}) = 0 \,,
 }
where from now on we shall denote the radial direction by $x_0$ or
$y_0$ and $\vec{x}$ is a vector with components along the $d$ remaining directions. The two linearly independent solutions to this equation are:
$\phi_1 = x_0^2 e^{-i \vec{k}\cdot\vec{x}} I_{\nu}(kx_0)$ and $\phi_2 = x_0^2 e^{-i \vec{k}\cdot \vec{x} } I_{-\nu} (kx_0)$ where $\nu = \sqrt{m^2 L^2 + {d^2 \over 4}}$. In the notation of our previous sections, the Green's function obeys the equation
 \eqn{GreenF}{
 (\square - m^2) G_E(x,y;f) =  \delta^{d+1} (x-y) \,,
}
where we remind ourselves that the delta function includes a ${1 \over
{\sqrt{g}}}$ in its definition. The right hand side is zero for $x_0
\neq y_0$, so we have
\eqn{Greenansatz}{\eqalign{
 G_E(x,y) &= A_1 \phi_1(x) + A_2 \phi_2(x)  \qquad  \hbox{for} \qquad x_0 < y_0 \, \cr
        &= B_1 \phi_1(x) + B_2 \phi_2(x) \qquad \hbox{for} \qquad x_0
> y_0 \,.  }} The boundary behavior of the scalar we're interested in
is: $\phi(x_0, \vec{x}) = f \beta(\vec{x}) x_0^{{d \over 2} + \nu} +
\beta(\vec{x}) x_0^{{d \over 2} - \nu}$. We choose our $\phi_1$ and
$\phi_2$ so that they have the right boundary behavior and also
require that the Green's function not diverge in the bulk (large
values of the radial coordinate $x_0$) for two non-coincident
points. One convenient choice of $\phi_1$ and $\phi_2$ is:
\eqn{basis}{\eqalign{
 \phi_1 &= x_0^{d \over 2} e^{-i {\vec{k}}.{\vec{x}}} \left( I_{\nu}(kx_0) + f ({2 \over k})^{2n} {{\Gamma(1 + \nu)} \over {\Gamma(1 - \nu)}} I_{\nu} (kx_0) \right) \, \qquad \hbox{and} \qquad 
 \phi_2 = x_0^{d \over 2} e^{-i {\vec{k}}.{\vec{x}}} K_{\nu} (k x_0) \,,
}} 
so that $\phi_1$ satisfies the boundary condition for small $x_0$
and $\phi_2$ is finite in the bulk. From the asymptotics of Bessel
functions, we see that $\phi_1$ diverges as $x_0 \to \infty$ and
$\phi_2$ diverges as $x_0 \to 0$. This forces us to set $A_2 = B_1 =
0$ in \Greenansatz. The remaining two constants are determined by
integrating \GreenF\ twice which gives us two conditions: (i) the
Green's function is continuous at $x_0 = y_0$ , and (ii) its radial
derivative has a jump discontinuity of $1 \over x_0^{d-1}$ at $x_0 =
y_0$. This yields \eqn{Greenconst}{\eqalign{
 A_1 &= {{\phi_2 (y_0)} \over {{\cal W} [\phi_1 (y_0), \phi_2 (y_0)]}} \, \qquad
 B_2 = {{\phi_1 (y_0)} \over {{\cal W} [\phi_1 (y_0), \phi_2 (y_0)]}} \,, 
}}
where ${\cal W} [\phi_1 (y_0), \phi_2 (y_0)]$ is the Wronskian. For our choice of $\phi_1$ and $\phi_2$ the Wronskian is:
\eqn{Wronskian}{
 {\cal W} [\phi_1 (y_0), \phi_2 (y_0)] = - {{\Gamma(1 - \nu) + f ({2 \over k})^{2 \nu} \Gamma(1 + \nu)} \over {\Gamma(1 - \nu)}} \left( {L \over {y_0}} \right)^{d-1}\,,
}
so combining \Greenansatz, \Greenconst, and \Wronskian\ we obtain the Green's function:
\eqn{EuclGreen}{
 G_E(x,y;f) = - \int {{d \kappa_E d^{d-1} k} \over {(2 \pi)^d}} {e^{-i \vec{k} \cdot (\vec{x} - \vec{y})} {(x_0 y_0)^{d \over 2} K_{\nu}(ky_0)} \over {\left( 1 + ({2 \over k})^{2 \nu} f {{\Gamma(1 + \nu)} \over {\Gamma(1 - \nu)}} \right)} L^{d-1} }  \left[ I_{- \nu}(k x_0) + f {{\Gamma (1 + \nu)} \over {\Gamma(1 - \nu)}} ({2 \over k})^{2 \nu} I_{\nu}(kx_0) \right] 
}
for $x_0 < y_0$ and a similar expression for $x_0 > y_0$. In the above equation, $\kappa_E$ is the temporal component of momentum. Finally, we Wick rotate this component $ \kappa_E = i k$ to get the Green's function in Minkowski signature:
\eqn{FullGreen}{
i G(x,y;f) =  \int {{d^d k} \over {(2 \pi)^d}} {e^{-i \vec{k} \cdot (\vec{x} - \vec{y})} {(x_0 y_0)^{d \over 2} K_{\nu}(ky_0)} \over {\left( 1 + ({2 \over k})^{2 \nu} f {{\Gamma(1 + \nu)} \over {\Gamma(1 - \nu)}} \right)} L^{d-1} }  \left[ I_{- \nu}(k x_0) + f {{\Gamma (1 + \nu)} \over {\Gamma(1 - \nu)}} ({2 \over k})^{2 \nu} I_{\nu}(kx_0) \right] 
}

The integral for general values of $f$, $d$ and $\nu$ is hard. For
$f=0$ and $f= \infty$ it can be evaluated and the result is an
expression which is related to \GreensFunct\ by a quadratic
hypergeometric transformation \cite{Muck}. A little bit more
can be said about the radial dependence of the one-loop vacuum
energy. This latter quantity depends on the Green's function for
coincident points $G(x,x;f)$. We saw before that this divergent
quantity was best handled by subtracting out $G(x,x;0)$. The result is
then finite:
 \eqn{DeltaG}{
 i\left[ G(x,x;f) - G(x,x;0) \right] = - {{1} \over {2^{d-2} \pi^{{d \over 2}} L^{d-1}  \Gamma (\nu) \Gamma(1 - \nu) \Gamma({d \over 2})}} \int_0^{\infty} d \tilde{k} \tilde{k}^{d-1} {\tilde{f} \over {\tilde{k}^{2 \nu} + \tilde{f} }} [K_{\nu} (\tilde{k})]^2 \,,  
}
where $\tilde{f} = 2^{2 \nu} {{\Gamma(1+ \nu)} \over {\Gamma(1 - \nu)}} f x_0^{2 \nu}$ and $\tilde{k} = k x_0$. Note that the excess vacuum energy depends on the radial coordinate $x_0$ in the particular combination $f x_0^{2 \nu}$. 

In order to make any further progress, one would need to first compute
the momentum integral and then integrate over $\nu$ to obtain the
vacuum energy. We argued earlier that $V(x;m_{BF}^2,f) - V(x;m_{BF}^2,0) = 0$ for all values of $f$, so using \FinalVDiff\  and \DeltaG\ we have:
 \eqn{HardInt}{\eqalign{
   V(x;m^2,f) - V(x;m^2,0) &= {i \over 2} \int_{m_{BF}^2}^{m_0^2} \ d \tilde{m}^2 \left[ G(x,x;f) - G(x,x;0) \right] \, \cr
  &= {i \over 2} \int_{0}^{\nu} \ {d \tilde{\nu}^2 \over L^2} \left[ G(x,x;f) - G(x,x;0) \right] \, \cr
  &= - {1 \over {2^{d-2} \pi^{{d \over 2}} \Gamma({d \over 2}) 
      L^{d+1}}} \int_0^{\nu} \ d \tilde{\nu} { \tilde{\nu} \over {\Gamma(\tilde{\nu}) \Gamma(1 - \tilde{\nu})}} 
      \int_0^{\infty} \ d \tilde{k} {{\tilde{k}^{d-1} \tilde{f}} \over {\tilde{k} + \tilde{f}}} \left[ K_{\nu}(\tilde{k}) \right]^2  \,,
 }}
where we remind ourselves that $\tilde{f} = 2^{2 \nu} {{\Gamma(1+
\nu)} \over {\Gamma(1 - \nu)}} f x_0^{2 \nu}$.  The double integral is
difficult to perform explicitly.  However, it is not hard to show from
\HardInt\ that $V(x;m^2,f)$ decreases monotonically as $f$ increases
from $0$ to $\infty$. To see this we note that the integrand depends on $x_0$ only through $\tilde{f}$ and since the integrand is a monotonic function of $\tilde{f}$, clearly $V(x;m^2,f)$ decreases monotonically with increasing $f$.

\section{Conclusions}
\label{Conclusions}

The upshot of section~\ref{Endpoints} was an evaluation of the change
in the one-loop self-energy, $V_+ - V_-$, between the IR and UV
endpoints of a holographic RG flow.  We would now like to convert this
into a change in the central charge of the dual field theory.

In \cite{HennSken}, the central charge was obtained by holographically
computing the Weyl anomaly: on the field theory side,
 \eqn{WeylDef}{
  \delta W[g_{\mu\nu}] = 
   {1 \over 2} \int d^4 x \sqrt{g} \, \omega 
    \langle T^\mu_\mu \rangle
 }
upon a conformal variation $g_{\mu\nu} \to e^{2\omega} g_{\mu\nu}$,
where $W$ is the generating functional for connected Green's
functions.  At the one-loop level, the prescription of
\cite{GKP,WittenAdS} asserts that $W$ is the classical supergravity
action.  The exact statement is that the partition functions of string
theory and gauge theory coincide (subjected to boundary conditions and
source terms in the usual way).  In the calculation of
\cite{HennSken}, the supergravity action integral is evaluated with a
radial cutoff, where the choice of radius amounts to a choice of
metric within a conformal class.  The supergravity lagrangian
evaluates to a constant in AdS, and the central charge is proportional
to this constant.\footnote{{\it A priori}, one might worry that
boundary terms in the supergravity action also contribute to the
central charge.  That this does not happen depends on the
circumstance, noted in \cite{HennSken}, that the only log-divergent
terms in the supergravity calculation arise from the integral of the
bulk action.}  All that we need to do in order to correct the central
charge computation at one loop is to ask by how much the
one-loop-corrected lagrangian differs from the tree-level lagrangian,
when evaluated in AdS.  The tree level and one-loop lagrangians will
stand in the same ratio as the leading large $N$ central charge and
its $1/N$-corrected counterpart.

The tree level lagrangian is 
 \eqn{LTree}{
  \sqrt{g}^{-1} {\cal L}_{\rm tree} = 
   {1 \over \kappa_{d+1}^2} (R-\Lambda_0) = 
   -{2d \over \kappa_{d+1}^2 L^2} \,.
 }  
The calculation indicated by the discussion in the previous paragraph
is
 \eqn{cTreeLoop}{
  {c_{\rm corrected} \over c_{\rm tree}} = 
   {{\cal L}_{\rm tree} + \delta {\cal L} \over
    {\cal L}_{\rm tree}} \,,
 }
where $\delta {\cal L} = -\sqrt{g} V$ is the one-loop correction to the
lagrangian that we computed in section~\ref{OneLoop}.  Because we are
only able to compute $V$ up to an additive constant that is
independent of boundary conditions, the only meaningful ratio that we
can form is
 \eqn{cfDep}{
  {c_+ \over c_-} = {{\cal L}_{\rm tree} - \sqrt{g} V_+ \over 
   {\cal L}_{\rm tree} - \sqrt{g} V_-} = 
   1 + {V_- - V_+ \over { \sqrt{g}^{-1} \cal L}_{\rm tree}} \,.
 }
so that 
 \eqn{Signc}{
 {{c_+ - c_-} \over c_-} =  (V_+ - V_-) \left({{\kappa^{2}_{d+1} L^2} \over {2d}} \right) \,.
 }
To check if $c_-$ is indeed greater than $c_+$, all that we have to
show is that $V_+ < V_-$. But that is exactly what we saw above.

As an example, in $AdS_5$, we obtain from \OddVdiff\ and \Signc\ the
result
 \eqn{FiveEx}{
  {c_+ - c_- \over c_-} = 
   {\kappa_5^2 \over 120 \pi^2 L^3} \left[ {(\Delta_- - 2)^3 \over 3} - 
    {(\Delta_- - 2)^5 \over 5}  \right] \,.
 }

One can go further and translate the function $V(z;m^2,f)-V_-$ into a
correction to the central charge whose scale dependence is monotonic.
It is not clear how well-defined such a function can be on the
supergravity side: because the bulk theory includes gravity, it has no
local observables.  Poetically, we would like to relate this to the
fact that renormalization group effects in field theory are
scheme-dependent---but it is difficult to make this precise.

It would be interesting to see how the construction discussed in this
paper might be realized as part of a compactification of string theory
to four dimensions, along the lines of \cite{VerlindeCY,Giddings:2001yu}.
One of the most interesting questions in that context is one that we
glossed over here: before considering the loop effects in
supergravity, one generally expects a moduli space of vacua, and this
statement probably translates into field theory terms as the existence
of a line of fixed points.  Mapping the lifting of moduli into field
theory terms might at least gain us a restatement of the moduli
problem in terms of the existence of isolated fixed points of the
renormalization group.

\section*{Acknowledgments}

This work was supported in part by DOE grant DE-FG02-91ER40671.  We
thank E.~Silverstein and E.~Witten for useful discussions.

\section*{Appendix}

In this appendix we shall sketch the derivation of \ReplaceV. Our
starting point is the familiar field theory result that the one-loop
effective potential is
 \eqn{VDisplay}{
  V(z;m^2,f) = - {i \over 2} {\rm tr} \log (- \square + m^2) \,.
 }
We shall denote the Klein-Gordon operator $(- \square + m^2)$ by
${\hat K}(m^2,f)$ and as an operator, it is related to our definition
of the Green's function \RepGEOM\ by ${\hat G}(m^2,f) = - [{\hat
K}(m^2,f)]^{-1}$. The representations of operators such as ${\hat
G}(m^2,f)$ in an orthonormal basis shall be denoted by the obvious
notation: $\langle z| {\hat G}(m^2,f) |z \rangle = G(z,z';m^2,f)$. In
terms of the Green's function, the effective potential is then
 \eqn{SimplerV}{
  V(z;m^2,f) = {i \over 2} \lim_{z' \to z} \log [-G(z,z';m^2,f)] \,.
 }
We shall use the Schwinger proper-time formalism to evaluate this. One needs two simple operator relations both of which follow from the relation between ${\hat G}(m^2,f)$ and ${\hat K}(m^2,f)$
 \eqn{GKRelate}{\eqalign{
  {\hat G}(m^2,f) &= - i \int_{0}^{\infty} e^{-is{\hat K}(m^2,f)} ds \, \cr
  \log [-{\hat G}(m^2,f)] &= \int_{0}^{\infty} {e^{- is{\hat K}(m^2,f)} \over {is}} ids  + \gamma \,,
 }}
where $\gamma$ is the Euler's constant. For the effective potential,
we see from \SimplerV\ that we need $\log[-{\hat G}(m^2,f)]$ which
differs by a factor of $is$ from the integral representation of ${\hat
G}(m^2,f)$ above.

To proceed any further we need the DeWitt-Schwinger representation of the Green's function (the reader is referred to \cite{BDBook} pg. 75 for a derivation)
\eqn{repG}{
 G(z,z';m^2,f) = - i {\sqrt{M(z,z')} \over {(4 \pi is)^{{d+1} \over
 2}}} \int_{0}^{\infty} i ds e^{-i m^2 s + {\eta(z,z') \over {2 i s}}}
 F(z,z';is) \,,
} 
where $\eta(z,z')$ is one-half the proper distance
 between the points $z$ and $z'$, and $M(z,z') = - {\rm det}
 [\partial_\mu \partial_\nu \eta(z,z')]$. For our purposes we shall
 just need to use the fact that the only place where the mass appears
 is in the exponent and integrating with respect to $m^2$ will bring
 down an extra factor of $is$ that we need. So integrating both sides
 of \repG\ between two arbitrary masses $m_1^2$ and $m_2^2$ and using
 \GKRelate\ we obtain \eqn{Diff}{
 \int_{{m_1}^2}^{{m_2}^2} d{\tilde m}^2 [-G(z,z';{\tilde m}^2,f)] = - \log [- G(z,z';m_2^2,f)] + \log [- G(z,z';m_1^2,f)] \,.
}
In the usual treatment one chooses one of the masses to be infinite,
but as we explained in the main text, this cannot be done
here. Instead of integrating toward heavier masses, we integrate in
the opposite direction down to the Breitenlohner-Freedman
bound. Therefore, we set $m_1^2 = m_{{\rm BF}}^2$ and $m_2^2 = m^2$ in
\Diff\ and use (\ref{SimplerV}) to get 
\eqn{UseV}{
 V(z;m^2,f) = {i \over {2}} \lim_{z \to z'} \int_{m_{\rm BF}^2}^{m^2} d {\tilde m}^2 G(z,z';m^2,f)  + V(z;m_{\rm BF}^2,f) \,.
}

\bibliography{loop}
\bibliographystyle{ssg}

\end{document}